\documentclass[11pt]{article}%
\usepackage{amsmath,amssymb,amsthm,graphicx,slashed}
\usepackage{amsmath}
\usepackage{amsfonts}
\usepackage{amssymb}
\usepackage{tikz,pgfplots}
\usepackage{tikz-feynman}
\usepackage{graphicx}%
\providecommand{\U}[1]{\protect\rule{.1in}{.1in}}

\DeclareMathOperator{\tr}{Tr}
\def\U{\mathcal{U}}
\begin{document}

\title{Spectral action in matrix form}

\author
      {Ali H. Chamseddine$^{1}$, John Iliopoulos$^{2}$ and Walter D. van
Suijlekom$^{3}$
\\[4mm]
\parbox{.9\linewidth}{
  \centering
     $^{1}$Physics Department, American University of Beirut, Lebanon\\
           $^{2}$Laboratoire de Physique - ENS - CNRS, PSL Research University
             and Sorbonne Universités,\\ 24 rue Lhomond, F-75231 Paris Cedex 05, France \\ 
           $^{3}$Institute for Mathematics, Astrophysics and Particle Physics, Radboud
University Nijmegen, Heyendaalseweg 135, 6525 AJ Nijmegen, The Netherlands.}
\\[4mm]
\texttt{chams@aub.edu.lb, ilio@ens.fr, waltervs@math.ru.nl}
      }

\maketitle
\begin{abstract}
Quantization of the noncommutative geometric spectral action has so far been performed on the final component form
of the action where all traces over the Dirac matrices and symmetry algebra are carried out. In this work, in order
to preserve the noncommutative geometric structure of the formalism, we derive the quantization rules for propagators and vertices
in matrix form. We show that the results in the case of a product of a four-dimensional Euclidean manifold by a finite
space, could be cast in the form of that of a Yang-Mills theory. We illustrate the procedure for the toy electroweak model.
 \end{abstract}

\section{Introduction}
Active research during the last decade on the noncommutative nature of
space-time has enhanced our understanding of the fundamental particles in
nature and their interactions (we refer to \cite{CCM07,john,Liz18,AliWalter,Sui14} for an overview of the literature). The emergent picture is that of a space-time
which is a product of continuous four-dimensional manifold with a
finite space, determined uniquely from very few physical requirements \cite{CC07b}. In particular, when the order one condition is imposed on the Dirac operator, we obtained all the spectrum of the Standard Model, in addition to the
right-handed neutrinos and a scalar singlet whose vev generate Majorana mass
to the right-handed neutrinos The fermionic action is of the Dirac type while
the classical bosonic action is determined from the spectral action principle.
As all gauge and Higgs fields are unified within the same Dirac operator, this
resulted in certain relations between the three gauge couplings and also the
Higgs coupling. These relations were taken to hold only at some unification
scale and the action considered as an effective action from the Wilsonian
point of view ({\em cf.} \cite{AliWalter} and references therein). The renormalization program was then carried out according to the rules of perturbative quantum field theory to the final derived form of the action. One immediately recognizes that following this method a vast amount of information
is lost because the action is obtained after taking all traces
of the matrices of the Clifford and symmetry algebras. Moreover, looking at the
component form of the propagators of all the dynamical fields it is not
possible to see the unified picture they were derived from. In particular, it is known \cite{JI} that in the traditional framework of the Standard Model relations among coupling constants are not stable under renormalization. In contrast, in our formulation all the fermionic fields are assembled in one spinor, acted on with a Dirac
operator $D,$ which is a $384\times384$ matrix implying that the fermionic
propagator $D^{-1}$ is also a matrix of the same dimension. In order not to
lose the noncommutative nature of the space at the quantum level, especially
of the finite part, it is important to derive all propagators and vertices in
matrix form. In this paper we will achieve this by expressing the bosonic action in terms of matrix-valued fields, without taking the traces. For simplicity, and to avoid the complications of including the gravitational field, we will assume a flat four-dimensional Euclidean manifold. It is possible to generalize this work to include gravity by using results from our previous works on the spectral action (see \cite{AliWalter} for more details and references).

\section{Action in matrix form}

Consider a noncommutative space $\left(  \mathcal{A},\mathcal{H}%
,D,J,\gamma\right)  $ defined as the product of a four-dimensional manifold $M$ with the spectral data $\left(  C^{\infty
}\left(  M\right)  ,L^{2}\left(  S\right)  ,i\gamma^{\mu}\partial_{\mu
},C,\gamma_{5}\right)  $ where $C$ is the charge conjugation operator, times a
finite space $\left(  \mathcal{A}_{F},\mathcal{H}_{F},D_{F},J_{F},\gamma
_{F}\right)  $. \ The spectral data is given by $\left(  \mathcal{A}%
,\mathcal{H},D,J,\gamma\right)  $ where \cite{CC96}
\begin{align}
\mathcal{A}  &  =C^{\infty}\left(  M\right)  \otimes\mathcal{A}_{F} \nonumber \\
\mathcal{H}  &  =L^{2}\left(  S\right)  \otimes\mathcal{H}_{F} \nonumber\\
D  &  =i\gamma^{\mu}\partial_{\mu}\otimes1+\gamma_{5}\otimes D_{F}%
\label{Dirac}\\
J  &  =C\otimes J_{F}\nonumber\\
\gamma &  =\gamma_{5}\otimes\gamma_{F} \nonumber
\end{align}
The Dirac operator including inner fluctuations is
\[
D_{A}=i\gamma^{\mu}\partial_{\mu}+A\text{ }%
\]
where
\begin{equation*}
A=%
{\displaystyle\sum}
a\widehat{a}\left[  D,b\widehat{b}\right]  ,\qquad a,b\in\mathcal{A}%
,\qquad\widehat{a},\widehat{b}\in\mathcal{A}^{0}.%
\end{equation*}
and where $\mathcal{A}^{0}$ is the opposite algebra formed from elements
$Ja^{\ast}J^{-1},$ $a\in\mathcal{A}.$ The connection $A$, in the general case,
takes the form \cite{CCS15}
\begin{align*}
A  &  =A_{(1)}+JA_{(1)}J^{-1}+A_{(2)}\\
A_{(1)}  &  =%
{\displaystyle\sum}
a\left[  D,b\right]  ,\qquad A_{(2)}=%
{\displaystyle\sum}
\widehat{a}\left[  A_{(1)},\widehat{b}\right]
\end{align*}
The Dirac matrices $\gamma^{\mu}$ satisfy $\left\{  \gamma^{\mu},\gamma^{\nu
}\right\}  =2\delta^{\mu\nu}$ and $\gamma^{\mu\ast}=\gamma^{\mu}.$ Squaring
$D$ we get
\[
D^{2}=-\left(  \partial^{\mu}\partial_{\mu}+\mathcal{A}^{\mu}\partial_{\mu
}+\mathcal{B}\right)
\]
where
\begin{align*}
\mathcal{A}^{\mu}  &  =-i\left\{  \gamma^{\mu},A\right\} \\
\mathcal{B}  &  \mathcal{=-}\left(  i\gamma^{\mu}\partial_{\mu}A+A^{2}\right)
\end{align*}
then $D^{2}$ can be written in the form $-\nabla^{\mu}\nabla_{\mu}-E$ where
$\nabla_{\mu}=\partial_{\mu}+\omega_{\mu}$ and
\begin{align*}
\omega_{\mu}  &  =\frac{1}{2}\mathcal{A}_{\mu}=-\frac{1}{2}i\left\{
\gamma_{\mu},A\right\} \\
E  &  =\mathcal{B-\partial}^{\mu}\omega_{\mu}-\omega^{\mu}\omega_{\mu}.%
\end{align*}
Then the covariant quantities are $E$
\[
E=\frac{1}{2}i\left[  \partial_{\mu}A,\gamma^{\mu}\right]  +\frac{1}{4}\left(
\gamma_{\mu}A\gamma^{\mu}A+A\gamma_{\mu}A\gamma^{\mu}+\gamma_{\mu}A^{2}%
\gamma^{\mu}\right)
\]
and the curvature $\Omega_{\mu\nu}$%
\begin{align*}
\Omega_{\mu\nu}  &  =\partial_{\mu}\omega_{\nu}-\partial_{\nu}\omega_{\mu
}+\left[  \omega_{\mu},\omega_{\nu}\right] \\
&  =-\frac{1}{2}i\left\{  \gamma_{\nu},\partial_{\mu}A\right\}  +\frac{1}%
{2}i\left\{  \gamma_{\mu},\partial_{\nu}A\right\}  -\frac{1}{4}\left[
\left\{  \gamma_{\mu},A\right\}  ,\left\{  \gamma_{\nu},A\right\}  \right].
\end{align*}
We now use the Gilkey formulas used in all our previous spectral action
calculations (in flat space)
\begin{align*}
a_{0}  &  =\frac{1}{16\pi^{2}}%
{\displaystyle\int}
d^{4}x\tr\left(  1\right) \\
a_{2}  &  =\frac{1}{16\pi^{2}}%
{\displaystyle\int}
d^{4}x\tr\left(  E\right) \\
a_{4}  &  =\frac{1}{16\pi^{2}}\frac{1}{12}%
{\displaystyle\int}
d^{4}x\tr\left(  \Omega_{\mu\nu}\Omega^{\mu\nu}+6E^{2}\right)
\end{align*}
where the trace $\tr$ is over the Clifford algebra and matrix algebra of $A.$
The spectral action $f\left(  D^{2}\right)  $ then gives \cite{CC96}
\[
I=\Lambda^{4}f_{4}a_{0}+\Lambda^{2}f_{2}a_{2}+f_{0}a_{4}%
\]
where
\[
f_{4}=%
{\displaystyle\int\limits_{0}^{\infty}}
uf\left(  u\right)  du,\qquad f_{2}=%
{\displaystyle\int\limits_{0}^{\infty}}
f\left(  u\right)  du,\qquad f_{0}=f\left(  0\right)
\]
We will use a cutoff function and truncate all higher order terms.

The problem to express $I$ in terms of $A$ now becomes trivial but tedious, eventually resulting in
\begin{align*}
\tr\left(  E\right)   &  =\frac{1}{2}\tr\left(  2A^{2}+\gamma_{\mu}%
A\gamma^{\mu}A\right) \\
\tr \left(  \Omega_{\mu\nu}\Omega^{\mu\nu}+6E^{2}\right)   &  =- \tr\left(
\gamma_{\nu}\partial_{\mu}A\gamma^{\nu}\partial^{\mu}A+ 2\gamma_{\mu}%
\partial^{\mu}A\gamma_{\nu}\partial^{\nu}A\right) \\
&  \qquad+2 i\tr\left(  \partial_{\mu}A\gamma^{\mu}A\gamma_{\nu}A\gamma^{\nu
}-\gamma_{\mu}\partial^{\mu}A\gamma_{\nu}A\gamma^{\nu}A\right) \\
&  \qquad+\frac{1}{4}\tr\left(  \gamma_{\mu}A\gamma_{\nu}A\gamma^{\mu}%
A\gamma^{\nu}A+2\gamma_{\mu}A\gamma^{\mu}A\gamma_{\nu}A\gamma^{\nu}A\right)
\end{align*}
All contractions of the Dirac gamma matrices cancel. There is one set of cubic terms
that do not vanish, but combine to form a total divergence%
\[
- i \partial_{\mu}\tr\left(  \gamma^{\mu\nu}\left(  A^{2}\gamma_{\nu}%
A-A\gamma_{\nu}A^{2}\right)  \right)
\]
Thus the spectral action is given by
\begin{align}
I  &  =\frac{1}{32\pi^{2}}%
{\displaystyle\int}
d^{4}x\bigg( 2\Lambda^{4} f_{4}\tr\left(  1\right)  +\Lambda^{2}
f_{2}\tr\left(  2A^{2}+\gamma_{\mu}A\gamma^{\mu}A\right) \nonumber\\
&  \qquad\qquad-\frac{f_{0}}{6}\tr\left(  \gamma_{\mu}\partial_{\nu}%
A\gamma^{\mu}\partial^{\nu}A+2\gamma_{\mu}\partial^{\mu}A\gamma_{\nu}%
\partial^{\nu}A\right) \nonumber\\
&  \qquad\qquad+\frac{i f_{0}}{3}\tr\left(  \partial_{\mu}A\gamma^{\mu}%
A\gamma_{\nu}A\gamma^{\nu}- \gamma_{\mu}\partial^{\mu}A\gamma_{\nu}%
A\gamma^{\nu}A\right) \nonumber\\
&  \qquad\qquad\left.  +\frac{f_{0}}{24}\tr\left(  \gamma_{\mu}A\gamma_{\nu
}A\gamma^{\mu}A\gamma^{\nu}A+2\gamma_{\mu}A\gamma^{\mu}A\gamma_{\nu}%
A\gamma^{\nu}A\right)  \right)  \label{eq:lagrangian}%
\end{align}
An alternative derivation of this Lagrangian could be obtained based on
the functional analytical results of \cite{Sui11,Sui11d}. We refrain from
including it here as this lies beyond the scope of the present paper. Instead,
we now turn our attention to gauge fixing and the Feynman rules that this
Lagrangian gives rise to.

\section{Gauge fixing and Feynman rules}
\label{sect:feynman}
Let us analyze the properties of $I$ with respect to gauge transformations. We already mentioned that the quantities $E$ and $\Omega_{\mu\nu}$ are gauge covariant. To see this we start with the gauge
transformations of $A$ given by \cite{CCS15}
\[
A\mapsto u^{\ast}Au+u^{\ast}\delta u,\qquad u^{\ast}u=1 =  u u^\ast,
\]
where $\delta u=i\gamma^{\nu}\partial_{\nu}u,$ and $u$ has the same matrix
structure as $A,$ then with simple but tedious algebra we can verify that
\begin{align*}
E  &  \mapsto u^{\ast}Eu\\
\Omega_{\mu\nu}  &  \mapsto u^{\ast}\Omega_{\mu\nu}u
\end{align*}
transform covariantly. The connection $\omega_{\mu}$ transforms as
\[
\omega_{\mu}\mapsto u^{\ast}\omega_{\mu}u+u^{\ast}\partial_{\mu}u
\]
Note that for product  of continuous and finite spaces we can write
\[
A=i\gamma^{\mu}B_{\mu}+\gamma_{5}\phi
\]
as follows from the form of the Dirac operator \eqref{Dirac}, then
\[
\omega_{\mu}=-\frac{1}{2}i\left\{  \gamma_{\mu},A\right\}  =B_{\mu}%
\]
showing in this case that the curvature $\Omega_{\mu\nu}$ only depends on the gauge fields $B_\mu$.
Thus we must add to the action a gauge fixing term. Let
\[
\mathcal{G}=\partial_{\mu}\omega^{\mu}=-\frac{1}{2}i\left\{  \gamma^{\mu
},\partial_{\mu}A\right\}
\]
then we add to the action the term
\[
-\frac{1}{16\pi^{2}}\frac{f_{0}}{3\xi}%
{\displaystyle\int}
d^{4}x\tr\mathcal{G}^{2}=-\frac{1}{16\pi^{2}}\frac{f_{0}}{3\xi}%
{\displaystyle\int}
d^{4}x\tr\left(  \partial_{\mu}\omega^{\mu}\right)  ^{2}%
\]
The Feynman gauge corresponds to the choice $\xi=1.$ Adding a gauge fixing term
would then require introducing ghost fields $\overline{c}$ and $c$ with the
same matrix structure as $u.$ The form is determined from the infinitesimal
transformation of $\mathcal{G}$. Writing $u=1+i\alpha$ where $\alpha$ is
Hermitian we get that%
\[
\mathcal{G\mapsto}\left(  i\partial^{\mu}\partial_{\mu}\alpha+i\left[
\partial^{\mu}\omega_{\mu},\alpha\right]  +i\left[  \omega^{\mu},\partial
_{\mu}\alpha\right]  \right)
\]
and thus the ghost action is
\[
\int d^{4}x\tr\left(  \partial_{\mu}\overline{c}\nabla^{\mu}c\right)  =\int
d^{4}x\tr\left(  \partial_{\mu}\overline{c}\partial^{\mu}c-\frac{1}%
{2}i\partial_{\mu}\overline{c}\left[  \left\{  \gamma^{\mu},A\right\}
,c\right]  \right)  .
\]

\subsection{Feynman Rules}

We are now ready to give the Feynman rules.

The fermionic propagator corresponds to a massless fermion and is given by the
usual expression obtained from the (Euclidean) action of the form
$\Psi_{\alpha A}^{\ast}\left(  i \gamma^{\mu}\right)  _{\alpha}^{\beta
}\partial_{\mu}\Psi_{\beta A}$, that is to say,
\[
S_{\alpha A}^{\beta B}= - \frac{\left(  \gamma^{\mu}\right)  _{\alpha}^{\beta
}p_{\mu}\delta_{A}^{B}}{p^{2}}%
\]
To find the bosonic propagator we examine the following quadratic pieces of
the action%
\begin{align*}
&  \frac{f_{0} }{192\pi^{2}}\tr \left(  - \gamma_{\mu}\partial_{\nu}%
A\gamma^{\mu}\partial^{\nu}A- 2\gamma_{\mu}\partial^{\mu}A\gamma_{\nu}%
\partial^{\nu}A +\frac{1}{\xi} \left\{  \gamma_{\mu},\partial^{\mu}A\right\}
\left\{  \gamma_{\nu},\partial^{\nu}A\right\}  \right)
\end{align*}
Note that we do not include the quadratic ``mass'' terms $2A^{2}-\gamma_{\mu
}A\gamma_{\mu}A$ in the propagator; these will be taken into account using
vertices of order two, after which one can rely on the well-known identity,
\[
\frac{1}{P^{2}} + \frac{1}{P^{2}} M^{2} \frac{1}{P^{2}} + \cdots= \frac
{1}{P^{2}} \sum_{k \geq0}\left(  \frac{M^{2}}{P^{2}}\right)  ^{k} = \frac
{1}{P^{2}- M^{2}}
\]
thus re-assembling the pieces to arrive at a propagator that does include the
mass term.

Working in the Feynman gauge $\xi=1,$ the kinetic part of $A$ simplifies, but
it would not be easy to find the propagator. Indeed, one has to invert
expressions with gamma matrices. This can in fact be done, but the expression
obtained is rather involved. The problem at hand is to carry renormalization
program for models such as the Standard Model, and therefore it will be useful
to restrict ourselves to the case where%
\begin{align*}
A_{\alpha A}^{\beta B}  &  =\left(  i \gamma^{\mu}\right)  _{\alpha}^{\beta
}B_{\mu A}^{\quad B}+\left(  \gamma_{5}\right)  _{\alpha}^{\beta}\phi_{A}%
^{B}\\
&  \equiv B_{\alpha A}^{\beta B}+\Phi_{\alpha A}^{\beta B}%
\end{align*}
This should be anti-commuting with the grading $\gamma_{5} \otimes\gamma_{F}$
where $\gamma_{F}$ is a grading on the finite, matrix indices. Hence, with
respect to $\gamma_{F}$ the field $B_{\mu A}^{B}$ should be diagonal, while
the $\Phi$ should be off-diagonal.

The strategy is the following: express the quadratic expression in terms of
$B_{\mu}$ and $\phi,$ get the propagator for these fields, then plug back to
get the propagator for $A.$

First we have (in Feynman gauge)
\begin{align*}
&  \tr\left(  - \gamma_{\mu}\partial_{\nu}A\gamma^{\mu}\partial^{\nu}%
A-2\gamma^{\mu}\partial_{\mu}A\gamma^{\nu}\partial_{\nu}A+\left\{  \gamma
^{\mu},\partial_{\mu}A\right\}  \left\{  \gamma^{\nu},\partial_{\nu}A\right\}
\right) \\
&  =4\tr\left(  4B_{\mu}\partial^{2}B_{\mu}-6\phi\partial^{2}\phi\right)
\end{align*}
Hermiticity of $D$ and of $\gamma^{\mu}$ implies that
\[
B_{\mu}^{\ast}=-B_{\mu},\qquad\phi^{\ast}=\phi
\]
We then write
\[
B_{\mu A}^{\quad B}=iB_{\mu}^{i}\left(  T^{i}\right)  _{A}^{B},\qquad\phi
_{A}^{B}=\phi^{m}\left(  \lambda^{m}\right)  _{A}^{B}%
\]
where $\left(  T^{i}\right)  _{A}^{B}$ is a hermitian basis for the block
diagonal matrices, and $\left(  \lambda^{m}\right)  _{A}^{B}$ a hermitian
basis for the off-diagonal matrices satisfying
\[
\tr\left(  T^{i}T^{j}\right)  =\frac{1}{2}\delta^{ij},\qquad\tr\left(
\lambda^{m}\lambda^{n}\right)  =\frac{1}{2}\delta^{mn}%
\]
Thus the quadratic action becomes
\[
-\frac{f_{0}}{96\pi^{2}}\left(  4B_{\mu}^{i}\left(  \partial^{2}\right)
B_{\mu}^{i}+6\phi^{m}\left(  \partial^{2}
\right)  \phi^{m}\right).
\]
The propagators for $B_{\mu}^{i}$ and $\phi^{m}$ are
\begin{align*}
\left\langle B_{\mu}^{i}B_{\nu}^{j}\right\rangle  &  =\frac{12 \pi^{2}}{f_{0}%
}\delta^{ij}\delta_{\mu\nu}\frac{1}{p^{2}}\\
\left\langle \phi^{m}\phi^{n}\right\rangle  &  =\frac{8 \pi^{2}}{f_{0}}%
\delta^{mn}\frac{1}{p^{2} }%
\end{align*}
We now reassemble the pieces to get
\begin{align*}
\left\langle B_{\alpha A}^{\beta B}B_{\gamma C}^{\delta D}\right\rangle  &  =
-\frac{12 \pi^{2}}{f_{0}} \left(  \gamma^{\mu}\right)  _{\alpha}^{\beta
}\left(  \gamma_{\mu}\right)  _{\gamma}^{\delta}\left(  T^{i}\right)  _{A}%
^{B}\left(  T^{i}\right)  _{C}^{D}\frac{1}{p^{2}}\\
\left\langle \Phi_{\alpha A}^{\beta B}\Phi_{\gamma C}^{\delta D}\right\rangle
&  =\frac{8 \pi^{2}}{f_{0}}\left(  \gamma_{5}\right)  _{\alpha}^{\beta}\left(
\gamma_{5}\right)  _{\gamma}^{\delta}\left(  \lambda^{m}\right)  _{A}%
^{B}\left(  \lambda^{m}\right)  _{C}^{D}\frac{1}{p^{2} }%
\end{align*}
Thus the free propagator is
\begin{equation}
\left\langle A_{\alpha A}^{\beta B}A_{\gamma C}^{\delta D}\right\rangle =
\left(  \Gamma^{\tau I}\right)  _{\alpha A}^{\beta B} \left(  \Gamma_{\tau
I}\right)  _{\gamma C}^{\delta D} \frac{1}{p^{2}}%
\end{equation}
where we have defined
\[
\left(  \Gamma^{\tau I}\right)  _{\alpha A}^{\beta B} = \sqrt{\frac{12 \pi
^{2}}{f_{0}}}\delta^{\tau}_{\mu}\delta^{I}_{i} (i \gamma^{\mu})_{\alpha
}^{\beta}\left(  T^{i}\right)  _{A}^{B} + \sqrt{\frac{8 \pi^{2}}{f_{0}}}
\delta^{\tau}_{5} \delta^{I}_{m} (\gamma_{5})_{\alpha}^{\beta}\left(
\lambda^{m}\right)  _{A}^{B}
\]
Note that when an explicit fixed matrix structure is given, it is possible to
simplify the above formula somewhat further, using orthogonality results on
the matrix coefficients when summing over the $T^{i}$ and $\lambda^{m}$.

We have expressed the propagators appearing in the scale-invariant part of the
spectral action diagramatically using (ribbon) edges in Figure \ref{fig:edges}.


\begin{figure}[ptb]
\centering
\begin{tabular}
[c]{ll}%
\parbox{4cm}{
\begin{tikzpicture}
\begin{feynman}
\vertex(W);
\node at (W) [above right] {$\alpha A$};
\node at (W) [below right] {$\beta B$}; \vertex[right=of W] (C); \vertex[right=of C] (E); \node at (E) [above left] {$\gamma C$}; \node at (E) [below left] {$\delta D$}; \diagram* {(W) -- [double,thick] (E),} ;\end{feynman} \end{tikzpicture}} &
$\left\langle A_{\alpha A}^{\beta B}A_{\gamma C}^{\delta D}\right\rangle =
\left(  \Gamma^{\tau I}\right)  _{\alpha A}^{\beta B} \left(  \Gamma_{\tau
I}\right)  _{\gamma C}^{\delta D}\frac{1}{p^{2}}$\\[7mm]%
\parbox{4cm}{
\begin{tikzpicture}
\begin{feynman}
\vertex(W) {};
\vertex[right=of W] (C);
\vertex[right=of C] (E);
\diagram* {(W) -- [fermion,thick] (E),};
\node at (W) [below right] {$\alpha A$}; \node at (E) [below left] {$\beta B$}; \end{feynman} \end{tikzpicture}} &
$\langle\Psi_{\alpha A}^{\ast}\Psi_{\beta B}\rangle= - \left(  \gamma^{\mu
}\right)  _{\alpha}^{\beta}p_{\mu}\delta_{A}^{B} \frac1{p^{2}}$\\[5mm]%
\parbox{4cm}{\begin{tikzpicture}
\begin{feynman}
\vertex(W) {};
\vertex[right=of W] (C);
\vertex[right=of C] (E);
\diagram* {(W) -- [double,dashed,thick] (E),};
\node at (W) [below right] {$B$}; \node at (W) [above right] {$A$}; \node at (E) [below left] {$D$}; \node at (E) [above left] {$C$}; \end{feynman} \end{tikzpicture}} &
$\left\langle c_{A}^{\ast B}c_{C}^{D}\right\rangle = \left(  T^{i}\right)
_{A}^{B}\left(  T^{i}\right)  _{C}^{D}\frac{1}{p^{2}}$%
\end{tabular}
\caption{Feynman rules for the edges}%
\label{fig:edges}%
\end{figure}

\begin{figure}[ptb]
\centering
\begin{tabular}
[c]{ll}%
\parbox{4cm}{
\begin{tikzpicture}
\begin{feynman}
\vertex(W) {};
\vertex[right=of W] (C);
\vertex[right=of C] (E);
\node at (W) [above] {$\alpha A$};
\node at (W) [below] {$\beta B$}; \node at (E) [above] {$\gamma C$}; \node at (E) [below] {$\delta D$}; \diagram* {(W) -- [double,thick] (C), (C) -- [double,thick] (E),}; \filldraw (C) circle (3pt); \end{feynman} \end{tikzpicture}} &
$\frac{\Lambda^{2} f_{2}}{32 \pi^{2}} \left(  2 \delta_{\beta}^{\gamma}%
\delta_{B}^{C} \delta_{\delta}^{\alpha}\delta_{D}^{A} + \left(  \gamma_{\mu
}\right)  _{\beta}^{\gamma}\delta_{B}^{C} \left(  \gamma^{\mu}\right)
_{\delta}^{\alpha}\delta_{D}^{A} \right)  $\\[1cm]%
\parbox{4cm}{
\begin{tikzpicture}
\begin{feynman}
\vertex(W) {};
\vertex[above=of W] (NW);
\vertex[below=of W] (SW);
\vertex[right=of W] (C);
\vertex[right=of C] (E);
\node at (NW) [above right] {$\alpha A$};
\node at (NW) [below left] {$\beta B$}; \node at (SW) [above left] {$\gamma C$}; \node at (SW) [below right] {$\delta D$}; \node at (E) [above] {$\tau F$}; \node at (E) [below] {$\eta E$}; \diagram* {(SW) -- [double,thick] (C), (NW) -- [double,thick,momentum=\(p\)] (C), (C) -- [double,thick] (E),} ;\end{feynman} \end{tikzpicture}} &
$\frac{f_{0}}{96 \pi^{2}}p_{\mu}\delta_{B}^{C}\delta_{D}^{E}\delta_{F}%
^{A}\left(  \gamma_{\nu}\right)  _{\delta}^{\eta}\left(  \left(  \gamma^{\nu
}\right)  _{\beta}^{\gamma}\left(  \gamma^{\mu}\right)  _{\tau}^{\alpha
}-\left(  \gamma^{\mu}\right)  _{\beta}^{\gamma}\left(  \gamma^{\nu}\right)
_{\tau}^{\alpha}\right)  $\\[2cm]%
\parbox{4cm}{
\begin{tikzpicture}
\begin{feynman}
\vertex(W) {};
\vertex[above=of W] (NW);
\vertex[below=of W] (SW);
\vertex[right=of W] (C);
\vertex[right=of C] (E);
\vertex[above=of E] (NE);
\vertex[below=of E] (SE);
\node at (NW) [above right] {$\alpha A$};
\node at (NW) [below left] {$\beta B$}; \node at (SW) [above left] {$\gamma C$}; \node at (SW) [below right] {$\delta D$}; \node at (SE) [above right] {$\tau F$}; \node at (SE) [below left] {$\eta E$}; \node at (NE) [below right] {$\kappa G$}; \node at (NE) [above left] {$\lambda H$}; \diagram* {(SW) -- [double,thick] (C), (NW) -- [double,thick] (C), (C) -- [double,thick] (SE), (C) -- [double,thick] (NE)} ;\end{feynman} \end{tikzpicture}} &
$\frac{f_{0}}{768 \pi^{2}}\left(  \delta_{B}^{C}\delta_{D}^{E}\delta_{F}%
^{G}\delta_{H}^{A}\left(  \gamma_{\mu}\right)  _{\lambda}^{\alpha}\left(
\gamma_{\nu}\right)  _{\tau}^{\kappa}\left(  \left(  \gamma^{\nu}\right)
_{\beta}^{\gamma}\left(  \gamma^{\mu}\right)  _{\delta}^{\eta}+2\left(
\gamma^{\mu}\right)  _{\beta}^{\gamma}\left(  \gamma^{\nu}\right)  _{\delta
}^{\eta}\right)  \right)  $\\[2cm]%
\parbox{4cm}{\begin{tikzpicture}
\begin{feynman}
\vertex(W) {};
\vertex[right=of W] (C);
\vertex[right=of C] (E);
\vertex[above=of W] (NW);
\vertex[below=of W] (SW);
\node at (SW) [above left] {$\beta B$};
\node at (NW) [below left] {$\alpha A$}; \node at (E) [below] {$\delta D$}; \node at (E) [above] {$\gamma C$}; \diagram* {(E) -- [double,thick] (C), (NW) -- [fermion,thick] (C), (C) -- [fermion,thick] (SW),} ;\end{feynman} \end{tikzpicture}} &
$\delta^{\gamma}_{\alpha}\delta^{\beta}_{\delta}\delta_{A}^{C} \delta_{D}^{B}%
$\\[2cm]%
\parbox{4cm}{
\begin{tikzpicture}
\begin{feynman}
\vertex(W) {};
\vertex[right=of W] (C);
\vertex[right=of C] (E);
\vertex[above=of W] (NW);
\vertex[below=of W] (SW);
\node at (NW) [above right] {$A$};
\node at (NW) [below left] {$B$}; \node at (SW) [above left] {$C$}; \node at (SW) [below right] {$D$}; \node at (E) [above] {$\beta F$}; \node at (E) [below] {$\alpha E$}; \diagram* {(E) -- [double,thick] (C), (SW) -- [double, dashed,thick] (C), (NW) -- [double, dashed,thick,momentum=\(p\)] (C),} ;\end{feynman} \end{tikzpicture}} &
$p_{\mu} \left(  \gamma^{\mu} \right)  _{\beta}^{\alpha} \left(  \delta
_{F}^{C}\delta_{B}^{E}\delta_{D}^{A}-\delta_{B}^{C} \delta_{D}^{E}\delta
_{F}^{A}\right)  $%
\end{tabular}
\caption{Feynman rules for the vertices}%
\label{fig:vertices}%
\end{figure}
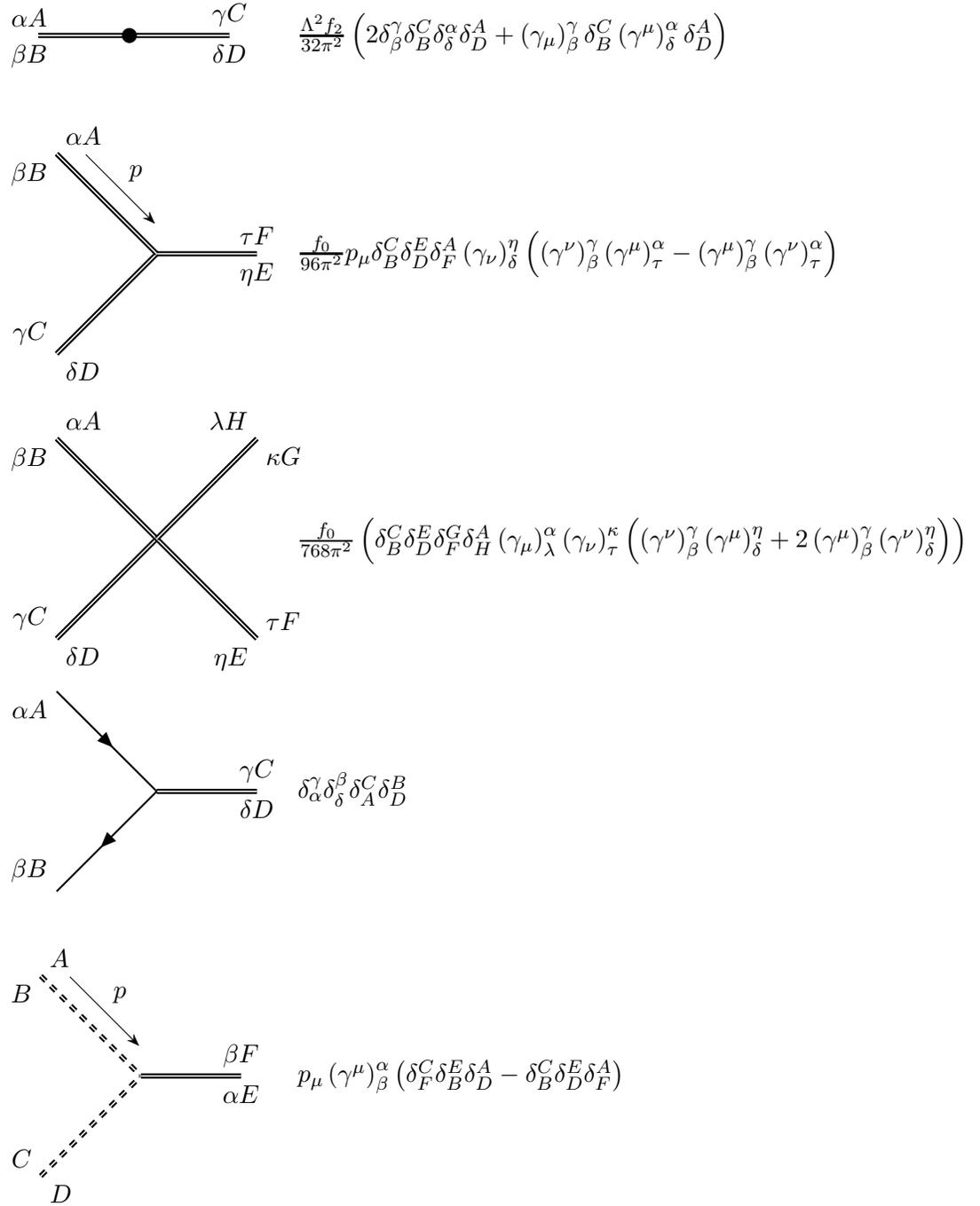\bigskip

Next we look at the vertices. We will start with the (negative) mass terms
that we will include as vertices of valence two. The relevant term is
\[
\frac{f_{2}}{32\pi^{2}}\tr\left(  2A^{2}+\gamma_{\mu}A\gamma^{\mu}A\right)
\]
which we may write in the form
\[
2 A_{\alpha A}^{\beta B} \delta_{\beta}^{\gamma}\delta_{B}^{C} A_{\gamma
C}^{\delta D}\delta_{\delta}^{\alpha}\delta_{D}^{A} + A_{\alpha A}^{\beta
B}\left(  \gamma_{\mu}\right)  _{\beta}^{\gamma}\delta_{B}^{C} A_{\gamma
C}^{\delta D}\left(  \gamma^{\mu}\right)  _{\delta}^{\alpha}\delta_{D}^{A}
\]
so that the contribution to the vertex is
\begin{equation}
V^{\alpha A \gamma C}_{\beta B \delta D} = \frac{\Lambda^{2} f_{2}}{32 \pi
^{2}} \left(  2 \delta_{\beta}^{\gamma}\delta_{B}^{C} \delta_{\delta}^{\alpha
}\delta_{D}^{A} + \left(  \gamma_{\mu}\right)  _{\beta}^{\gamma}\delta_{B}^{C}
\left(  \gamma^{\mu}\right)  _{\delta}^{\alpha}\delta_{D}^{A} \right)
\end{equation}
The first cubic contribution in the spectral action is
\[
- i\tr\left(  \partial_{\mu}A\gamma^{\mu}A\gamma_{\nu}A\gamma^{\nu}\right)
\]
and can be written in the form%
\[
p_{\mu}A_{\alpha A}^{\beta B}\left(  \gamma^{\mu}\right)  _{\beta}^{\gamma
}\delta_{B}^{C}A_{\gamma C}^{\delta D}\left(  \gamma_{\nu}\right)  _{\delta
}^{\eta}\delta_{D}^{E}A_{\eta E}^{\tau F}\left(  \gamma^{\nu}\right)  _{\tau
}^{\alpha}\delta_{F}^{A}%
\]
which shows that the contribution to the vertex is%
\[
p_{\mu}\left(  \gamma^{\mu}\right)  _{\beta}^{\gamma}\delta_{B}^{C}\left(
\gamma_{\nu}\right)  _{\delta}^{\eta}\delta_{D}^{E}\left(  \gamma^{\nu
}\right)  _{\tau}^{\alpha}\delta_{F}^{A}%
\]
The second cubic term is
\[
i \tr\left(  \gamma_{\mu}\partial^{\mu}A\gamma_{\nu}A\gamma^{\nu}A\right)
\]
We write it in the form%
\[
- p^{\mu}A_{\alpha A}^{\beta B}\left(  \gamma_{\nu}\right)  _{\beta}^{\gamma
}\delta_{B}^{C}A_{\gamma C}^{\delta D}\left(  \gamma^{\nu}\right)  _{\delta
}^{\eta}\delta_{D}^{E}A_{\eta E}^{\tau F}\left(  \gamma_{\mu}\right)  _{\tau
}^{\alpha}\delta_{F}^{A}%
\]
so that the contribution to the vertex is
\[
- p^{\mu}\left(  \gamma_{\nu}\right)  _{\beta}^{\gamma}\delta_{B}^{C}\left(
\gamma^{\nu}\right)  _{\delta}^{\eta}\delta_{D}^{E}\left(  \gamma_{\mu
}\right)  _{\tau}^{\alpha}\delta_{F}^{A}%
\]

Thus the cubic vertex is
\begin{equation}
V_{\beta B\delta D\tau F}^{\alpha A\gamma C\eta E}=\frac{f_{0}}{96 \pi^{2}%
}p_{\mu}\delta_{B}^{C}\delta_{D}^{E}\delta_{F}^{A}\left(  \gamma_{\nu}\right)
_{\delta}^{\eta}\left(  \left(  \gamma^{\nu}\right)  _{\beta}^{\gamma}\left(
\gamma^{\mu}\right)  _{\tau}^{\alpha}-\left(  \gamma^{\mu}\right)  _{\beta
}^{\gamma}\left(  \gamma^{\nu}\right)  _{\tau}^{\alpha}\right)
\end{equation}
The first quartic term
\[
\tr\left(  \gamma_{\mu}A\gamma_{\nu}A\gamma^{\mu}A\gamma^{\nu}A\right)
\]
can be written as
\[
A_{\alpha A}^{\beta B}\delta_{B}^{C}\left(  \gamma_{\nu}\right)  _{\beta
}^{\gamma}A_{\gamma C}^{\delta D}\delta_{D}^{E}\left(  \gamma_{\mu}\right)
_{\delta}^{\eta}A_{\eta E}^{\tau F}\delta_{F}^{G}\left(  \gamma^{\nu}\right)
_{\tau}^{\kappa}A_{\kappa G}^{\lambda H}\delta_{H}^{A}\left(  \gamma^{\mu
}\right)  _{\lambda}^{\alpha}%
\]
so that the contribution to the vertex is
\[
\delta_{B}^{C}\left(  \gamma_{\nu}\right)  _{\beta}^{\gamma}\delta_{D}%
^{E}\left(  \gamma_{\mu}\right)  _{\delta}^{\eta}\delta_{F}^{G}\left(
\gamma^{\nu}\right)  _{\tau}^{\kappa}\delta_{H}^{A}\left(  \gamma^{\mu
}\right)  _{\lambda}^{\alpha}%
\]
The next quartic term is
\[
2\tr\left(  \gamma_{\mu}A\gamma^{\mu}A\gamma_{\nu}A\gamma^{\nu}A\right)
\]
which can be written as
\[
2A_{\alpha A}^{\beta B}\delta_{B}^{C}\left(  \gamma_{\mu}\right)  _{\beta
}^{\gamma}A_{\gamma C}^{\delta D}\delta_{D}^{E}\left(  \gamma_{\nu}\right)
_{\delta}^{\eta}A_{\eta E}^{\tau F}\delta_{F}^{G}\left(  \gamma^{\nu}\right)
_{\tau}^{\kappa}A_{\kappa G}^{\lambda H}\delta_{H}^{A}\left(  \gamma^{\mu
}\right)  _{\lambda}^{\alpha}%
\]
so that the contribution to the vertex is
\[
2\delta_{B}^{C}\left(  \gamma_{\mu}\right)  _{\beta}^{\gamma}\delta_{D}%
^{E}\left(  \gamma_{\nu}\right)  _{\delta}^{\eta}\delta_{F}^{G}\left(
\gamma^{\nu}\right)  _{\tau}^{\kappa}\delta_{H}^{A}\left(  \gamma^{\mu
}\right)  _{\lambda}^{\alpha}%
\]
The total quartic vertex is then
\begin{equation}
V_{\beta B\delta D\tau F\lambda H}^{\alpha A\gamma C\eta E\kappa G}%
=\frac{f_{0}}{768 \pi^{2}}\left(  \delta_{B}^{C}\delta_{D}^{E}\delta_{F}%
^{G}\delta_{H}^{A}\left(  \gamma_{\mu}\right)  _{\lambda}^{\alpha}\left(
\gamma_{\nu}\right)  _{\tau}^{\kappa}\left(  \left(  \gamma^{\nu}\right)
_{\beta}^{\gamma}\left(  \gamma^{\mu}\right)  _{\delta}^{\eta}+2\left(
\gamma^{\mu}\right)  _{\beta}^{\gamma}\left(  \gamma^{\nu}\right)  _{\delta
}^{\eta}\right)  \right)
\end{equation}
Now for the ghost part we have%
\[
\tr\left(  -\overline{c}\partial_{\mu}\partial_{\mu}c\right)  = - \overline
c^{i}\partial^{2}c^{i}%
\]
where we have written $c_{A}^{B}=c^{i}\left(  T^{i}\right)  _{A}^{B}$ so that
\[
\left\langle c^{\ast i}c^{j}\right\rangle =\delta^{ij}\frac{1}{p^{2}}.
\]
We conclude that the ghost propagator is
\begin{equation}
\left\langle c_{A}^{\ast B}c_{C}^{D}\right\rangle =\left(  T^{i}\right)
_{A}^{B}\left(  T^{i}\right)  _{C}^{D}\frac{1}{p^{2}}%
\end{equation}
The vertex ghost terms are
\[
- \frac12 i \tr \left(  \partial_{\mu}\overline c \left[  \left\{  \gamma
_{\mu},A\right\}  ,c\right]  \right)
\]
which can be written as $\frac14 p_{\mu}\overline c_{A}^{B}\left(  \left(
\gamma_{\mu}\right)  _{\beta}^{\alpha}A_{\alpha E}^{\beta F}\right)  c_{C}%
^{D}\left(  \delta_{F}^{C}\delta_{B}^{E}\delta_{D}^{A}-\delta_{B}^{C}
\delta_{D}^{E}\delta_{F}^{A}\right)  $. Hence the ghost vertex becomes
\begin{equation}
\left\langle \overline c_{A}^{B}A_{\alpha E}^{\beta F}c_{C}^{D}\right\rangle
=\frac14 p_{\mu} \left(  \gamma^{\mu} \right)  _{\beta}^{\alpha} \left(
\delta_{F}^{C}\delta_{B}^{E}\delta_{D}^{A}-\delta_{B}^{C} \delta_{D}^{E}%
\delta_{F}^{A}\right)
\end{equation}
We have expressed the interaction terms appearing in the scale-invariant part
of the spectral action diagramatically using ribbon graphs in Figure
\ref{fig:vertices}.

\section{The Electroweak toy model}

Having used a super-condensed notation for the propagators, it would be
helpful to consider a toy model that captures the essential points of the
noncommutative unification. We thus consider the simple electroweak model where the
fermions are in the representation
\begin{equation*}
\Psi=\left(
\begin{array}
[c]{c}%
\nu_{L}\\
e_{L}\\
e_{R}%
\end{array}
\right)
\end{equation*}
and where the algebra $\mathcal{A}=M_{2}\left(  \mathbb{C}\right)
\oplus\mathbb{H}.$ The connection $A$ in this case is (ignoring the
complication of introducing the reality operator $J$ ) given by%
\begin{equation*}
A=\left(
\begin{array}
[c]{cc}%
i\left(  \gamma^{\mu}\right)  _{\alpha}^{\beta}\left(  B_{\mu}\right)
_{a}^{b} & \left(  \gamma_{5}\right)  _{\alpha}^{\beta}H^{b}\\
\left(  \gamma_{5}\right)  _{\alpha}^{\beta}H_{a} & i\left(  \gamma^{\mu
}\right)  _{\alpha}^{\beta}\left(  B_{\mu}\right)  _{3}^{3}%
\end{array}
\right)  ,\quad a,b=1,2
\end{equation*}
with the constraint that $\mathrm{Tr}A=0$ which implies that $\left(  B_{\mu
}\right)  _{a}^{a}+\left(  B_{\mu}\right)  _{3}^{3}=0.$ Thus in this notation,
when we write
\begin{equation*}
B_{\mu A}^{\quad B}=B_{\mu}^{i}\left(  T^{i}\right)  _{A}^{B},\quad\phi
_{B}^{A}=\phi^{m}\left(  \lambda^{m}\right)  _{A}^{B}%
\end{equation*}
then
\begin{align*}
\left(  T^{i}\right)  _{A}^{B}  & =\left(
\begin{array}
[c]{cc}%
\frac{1}{2}\left(  \sigma^{i}\right)  _{a}^{b} & 0\\
0 & 0
\end{array}
\right)  ,\quad i=1,2,3,\quad A=a,3\\
\left(  T^{0}\right)  _{A}^{B}  & =\left(
\begin{array}
[c]{cc}%
-\frac{1}{2\sqrt{3}}\delta_{a}^{b} & 0\\
0 & \frac{1}{\sqrt{3}}%
\end{array}
\right) ,\quad i=0
\end{align*}
and
\begin{align*}
\left(  \lambda^{1}\right)  _{A}^{B}  & =\left(
\begin{array}
[c]{ccc}%
0 & 0 & 1\\
0 & 0 & 0\\
1 & 0 & 0
\end{array}
\right)  ,\quad\left(  \lambda^{2}\right)  _{A}^{B}=\left(
\begin{array}
[c]{ccc}%
0 & 0 & i\\
0 & 0 & 0\\
-i & 0 & 0
\end{array}
\right)  \\
\left(  \lambda^{3}\right)  _{A}^{B}  & =\left(
\begin{array}
[c]{ccc}%
0 & 0 & 0\\
0 & 0 & 1\\
0 & 1 & 0
\end{array}
\right)  ,\quad\left(  \lambda^{2}\right)  _{A}^{B}=\left(
\begin{array}
[c]{ccc}%
0 & 0 & 0\\
0 & 0 & i\\
0 & -i & 0
\end{array}
\right)
\end{align*}
so that $H^{1}=\frac{1}{2}\left(  \phi^{1}+i\phi^{2}\right)  ,\quad
H^{2}=\frac{1}{2}\left(  \phi^{3}+i\phi^{4}\right)  ,$ $H_{1}=\overline{H^{1}%
},$ $H_{2}=\overline{H^{2}}.$ In this notation, the Dirac action becomes
\begin{align*}
\left(  \Psi,D\Psi\right)    & =\overline{l}i\gamma^{\mu}\left(  \partial
_{\mu}+\frac{1}{2}B_{\mu}^{p}\sigma^{p}-\frac{1}{2\sqrt{3}}B_{\mu}^{0}\right)
l+\overline{e}_{R}i\gamma^{\mu}\left(  \partial_{\mu}+\frac{1}{\sqrt{3}}%
B_{\mu}^{0}\right)  e_{R}\\
& +\overline{l}\gamma_{5}He_{R}+\overline{e}_{R}\gamma_{5}\overline{H}l
\end{align*}
where $l=\left(
\begin{array}
[c]{c}%
\nu_{L}\\
e_{L}%
\end{array}
\right)  .$ Comparing with the $SU\left(  2\right)  \times U\left(  1\right)
$ Weinberg-Salam leptonic model we have
\begin{equation*}
B_{\mu}^{p}=gW_{\mu}^{p},\quad B_{\mu}^{0}=g^{\prime}\sqrt{3}B_{\mu},\quad
p=1,2,3
\end{equation*}
However, from the spectral action we have
\begin{align*}
& -\frac{f_{0}}{24\pi^{2}}\left(  B_{\mu}^{i}\partial^{2}B_{\mu}^{i}+\frac
{3}{2}\phi^{m}\partial^{2}\phi^{m}\right)  \\
& =-\frac{f_{0}g^{2}}{24\pi^{2}}\left(  W_{\mu}^{p}\partial^{2}W_{\mu}%
^{p}+\frac{3g^{^{\prime}2}}{g^{2}}B_{\mu}\partial^{2}B_{\mu}+\frac{6}{g^{2}%
}H_{a}\partial^{2}H^{a}\right)
\end{align*}
Normalizing of the vector kinetic terms then requires
\begin{equation*}
\frac{f_{0}g^{2}}{12\pi^{2}}=1,\quad g^{^{\prime}2}=\frac{1}{3}g^{2}%
\end{equation*}
and thus
\begin{equation}
\label{sintheta}
\sin^{2}\theta_{W}=\frac{g^{^{\prime}2}}{g^{2}+g^{^{\prime}2}}=\frac{1}{4}%
\end{equation}
Normalizing the $H$ field kinetic term requires
\begin{equation*}
H\rightarrow\frac{g}{\sqrt{6}}H
\end{equation*}
which implies that the electron mass term is
\begin{equation}
\label{emass}
m_{e}=g\sqrt{\frac{1}{6}}\left\langle H\right\rangle
\end{equation}
The quartic Higgs coupling is
\begin{equation*}
\frac{f_{0}}{4\pi^{2}}\left(  \overline{H}H\right)  ^{2}\rightarrow\frac
{f_{0}}{4\pi^{2}}\frac{g^{4}}{36}\left(  \overline{H}H\right)  ^{2}=\frac
{1}{12}g^{2}\left(  \overline{H}H\right)  ^{2}%
\end{equation*}
and thus the coupling constant is
\begin{equation*}
\lambda=\frac{g^{2}}{12}%
\end{equation*}
We also note that this implies for the Higgs mass term the relation
\begin{equation}
\label{Hmass}
-\frac{\Lambda^{2}f_{2}}{2\pi^{2}}\overline{H}H\rightarrow-\frac{\Lambda
^{2}f_{2}g^{2}}{12\pi^{2}}\overline{H}H
\end{equation}
From the physical point of view, relations such as (\ref{sintheta}), (\ref{emass}) and (\ref{Hmass}), constitute important consequences of the underlying geometric structure and we wish to investigate their stability under renormalization.

\section{Conclusions}

In this note, we have reconsidered the renormalization program of the perturbative quantum
field theory resulting from the spectral action of a noncommutative space
formed as a product of a continuous four-dimensional manifold times a
finite space. The basic fields are the Fermi fields defining the Hilbert
space and the bosonic fields represented as inner fluctuations of the Dirac
operator of the noncommutative space. The bosonic fields include both vector
gauge and scalar Higgs fields. The crucial observation is to realize that the
nature of the finite space determining the matrix structure of the
Dirac operator, and the unification of gauge interactions, gauge and Higgs
coupling constants follows from this property.

As a first step to preserve the matrix
structure at the quantum level, we have derived the spectral action in matrix
form, and determined all propagators and vertices corresponding to matrix-valued fields, both in the Clifford algebra of the Dirac operator and algebra
$\mathcal{A}$ of the noncommutative space. We have also fixed the gauge
symmetry and determined the matrix-valued ghost fields associated with this
symmetry. The Feynman diagrams of the noncommutative spectral action, are then
in one-to-one correspondence with that of a Yang-Mills non-abelian gauge
theory, with a direct mass term. The main difference is the appearance of the
space-time Dirac matrices $\gamma^{\mu}$ and $\gamma_{5}$ not only for
fermionic fields, but also for bosonic fields. The next step is then to carry
out the analysis of the renormalizability of the spectral action, along the
same lines as those of a Yang--Mills theory, by first determining all the
possible Feynman diagrams. The main complication to do this analysis is the
appearance of ribbon diagrams thus making the computation technically
challenging. Note, however, the striking similarity to the diagrams that appear in noncommutative quantum field theory, starting with \cite{GW}. 
In any case, it is very promising to have prepared the ground to
engage in the task of performing the full renormalizability analysis of the
spectral model. We do expect that the novel procedure where the matrix
structure is preserved, to shed light on the unification of the gauge and
Higgs interactions. To gain further insight into the quantization of the 
noncommutative spectral models, it would be useful  to develop, in particular, the BRST
quantization for the ghost sector.
\bigskip

\bigskip

\textbf{{\large {Acknowledgments}}}

We would like to thank Alain Connes for illuminating discussions. The work of A.H.C is supported
in part by the National Science Foundation Grant No. Phys-1518371 and Phys-5912998. He also thanks the Radboud Excellence Initiative for hosting him at Radboud University where part of this research was carried out.


\newcommand{\noopsort}[1]{}\def\cprime{$'$}
\begin{thebibliography}{99}

\bibitem{CC96}
A.~H. Chamseddine and A.~Connes.
\newblock Universal formula for noncommutative geometry actions: {U}nifications
  of gravity and the {S}tandard {M}odel.
\newblock {\em Phys. Rev. Lett.} 77 (1996)  4868--4871.

\bibitem{CC07b}
A.~H. Chamseddine and A.~Connes.
\newblock {Why the Standard Model}.
\newblock {\em J. Geom. Phys.} 58 (2008)  38--47.

\bibitem{CCM07}
A.~H. Chamseddine, A.~Connes, and M.~Marcolli.
\newblock Gravity and the {S}tandard {M}odel with neutrino mixing.
\newblock {\em Adv. Theor. Math. Phys.} 11 (2007)  991--1089.

\bibitem{CCS13b}
A.~H. Chamseddine, A.~Connes, and W.~D. van Suijlekom.
\newblock {Beyond the spectral Standard Model: Emergence of Pati-Salam
  unification}.
\newblock {\em JHEP} 1311 (2013)  132.

\bibitem{CCS15}
A.~H. Chamseddine, A.~Connes, and W.~D. van Suijlekom.
\newblock Grand unification in the spectral {P}ati-{S}alam model.
\newblock {\em JHEP} 11 (2015)  011.

\bibitem{AliWalter}
A.~H. Chamseddine and W.~D. van Suijlekom.
\newblock {A Survey of Spectral Models of Gravity Coupled to Matter}
\newblock {\em Advances in Noncommutative Geometry, pp 1-51, Editors A. Chamseddine et al, Springer 2020}

\bibitem{GW}
H. Grosse and R. Wulkenhaar.
\newblock {Renormalisation of $\phi^4$-theory on noncommutative $\mathbb R^4$ in the matrix base}.
\newblock {\em Comm. Math. Phys.} 256 (2005), 305-374. 

\bibitem{JI}
  J.Iliopoulos
  Can we predict the value of the Higgs mass?
\newblock {hep-ph/0603146}

\bibitem{john}
J.~Iliopoulos
\newblock {Gauge theories and noncommutative geometry}
\newblock {\em Proceedings, 6th International Conference on New Frontiers in Physics ICNFP 2017}

\bibitem{Liz18}
F.~Lizzi.
\newblock {Noncommutative Geometry and Particle Physics}.
\newblock {\em PoS} CORFU2017 (2018)  133.

\bibitem{Sui11}
W.~D. van Suijlekom.
\newblock Perturbations and operator trace functions.
\newblock {\em J. Funct. Anal.} 260 (2011)  2483--2496.

\bibitem{Sui11d}
W.~D. van Suijlekom.
\newblock {Renormalizability conditions for almost commutative manifolds}.
\newblock {\em Ann. H. Poincar\'e} 15 (2014)  985--1011.


\bibitem{Sui14}
W.~D. van Suijlekom.
\newblock {\em Noncommutative Geometry and Particle Physics}.
\newblock Springer, 2015.


\end{thebibliography}

\end{document}